\documentclass[%
 reprint,
 amsmath,amssymb,
 aps,
 pre,
 showkeys
]{revtex4-2}

\usepackage{graphicx}
\usepackage{dcolumn}
\usepackage{bm}
\usepackage{mathtools}
\usepackage[colorlinks=true,linkcolor=magenta,citecolor=magenta,breaklinks=true]{hyperref}
\usepackage{lipsum}
\usepackage{color}

\newcommand{\Tc}{T_\mathrm{c}}

\begin{document}

\title{Neural Network Approach to Scaling Analysis of Critical Phenomena}

\author{Ryosuke Yoneda}
\email{yoneda.ryosuke.o95@kyoto-u.jp}
\author{Kenji Harada}
\email{harada.kenji.8e@kyoto-u.ac.jp}
\affiliation{
    Graduate School of Informatics, Kyoto University, Kyoto 606-8501, Japan
}

\date{\today}

\begin{abstract}
    Determining the universality class of a system exhibiting critical phenomena is one of the central problems in physics. There are several methods to determine this universality class from data. As methods performing collapse plots onto scaling functions, polynomial regression, which is less accurate, and Gaussian process regression, which provides high accuracy and flexibility but is computationally heavy, have been proposed. In this paper, we propose a regression method using a neural network. The computational complexity is only linear in the number of data points. We demonstrate the proposed method for the finite-size scaling analysis of critical phenomena on the two-dimensional Ising model and bond percolation problem to confirm the performance. This method efficiently obtains the critical values with accuracy in both cases.
\end{abstract}

\keywords{Neural networks, Scaling analysis, Finite-size scaling, Critical phenomena}

\maketitle

\section{Introduction}
Critical phenomena have been a significant research topic in statistical mechanics for many years. The scaling behavior near a critical point plays a central role in critical phenomena because critical phenomena of different physical systems share it. We can identify the universality class of critical phenomena by the scaling exponents called critical exponents. The universality class only depends on the dimensionality, the symmetry, and the type of interaction between components, not the details of systems. The renormalization group supports such property of the universality class for equilibrium systems~\cite{PhysRevB.4.3174,PhysRevB.4.3184, wilson1974renormalization,wilson1983renormalization,goldenfeld2018lectures,cardy1996scaling}. Also, in the case of non-equilibrium, the universality of critical phenomena is extensively studied~\cite{henkel2008non}. Therefore, identifying the universality class of critical phenomena is theoretically and experimentally important.

Even in the case of theoretical models, it is difficult to derive the value of critical exponents analytically in general. Thus, it is important to numerically estimate the value of critical exponents from given data. In particular, we often use the finite-size scaling (FSS) law to determine the value of critical exponents from data of finite-size systems~\cite{cardy2012finite}. We have applied the FSS analysis to various critical phenomena from classical and quantum systems~\cite{binder1981finite,slevin1999corrections,wang2006high,harada2013possibility,otsuka2016universal,yoneda2020critical}. The physical quantity near a critical point in a finite-size system obeys the scaling law written as
\begin{align}
    A(T,L)=L^{-c_{2}}F[(T-\Tc)L^{c_{1}}],
    \label{eq:fss}
\end{align}
where $A(T, L)$ is a physical quantity at temperature $T$ in a finite-size system of which size is $L$. $\Tc$ is a critical temperature. The exponent $c_1$ and $c_2$ are critical exponents. Here $F[\cdot]$ is a scaling function. Unfortunately, we do not know the scaling function's form in advance. Thus, if we directly use the FSS law, we need to infer not only the value of critical temperature and exponents but also the scaling function itself from the given data. In order to avoid this problem, several methods have been proposed. For example, we use the Binder ratio to determine the effective critical temperature for each system size, and so on. However, in the case of hard problems, the amount of data is often limited in a narrow region near a critical point. Thus, using all data not only in the narrow region is useful to estimate critical values. Here, we consider the direct use of the FSS law by collapsing data onto a scaling function. The approach is called FSS in the following.

The classical way to perform FSS has been to assume that the scaling function is polynomial and then use the least-squares method to determine the critical exponents by fitting the scaling law to data. However, this method has the problem of determining the degree of the polynomial while preventing overfitting. It also requires high accuracy in a very narrow region near the critical point because the representation power of polynomials is poor. To deal with this problem, a Bayesian regression method~\cite{harada2011, harada2015} was proposed to infer the scaling function and critical exponents, in which the scaling function is a sample of the adjusted Gaussian process. In this method, we assume only the smoothness of a scaling function. Thus, we can use data to fitting in the broader range near a critical point. This type of regression method is called the Gaussian process (GP) regression in the machine-learning field and is widely used to analyze real data~\cite{williams2006gaussian}. The GP for FSS~\cite{harada2011}, called Bayesian scaling analysis (BSA) in the following, has been widely used to determine critical exponents and scaling functions for various critical phenomena because of its high accuracy~\cite{harada2013possibility,nasu2015thermodynamics,singh2016signatures,hesselmann2016thermal,otsuka2016universal,d2017new,horita2017upper,ferrari2017competition,iino2019detecting,lang2019quantum,nomura2021dirac,yoneda2020critical}. However, the computational cost of the GP is a problem because the computation of a likelihood gradient needs the inverse matrix of the covariance of GP. The size of the matrix is $N \times N$, where $N$ is the number of data points. The total cost of the GP is proportional to $N^3$. Thus, it is expensive when the size of the data increases.

This paper proposes the \underline{n}eural network approach to \underline{s}caling \underline{a}nalysis (NSA), a new FSS method using a neural network. The main idea is to model the scaling function by a neural network and train it with data. Because of the greater expressive power of a neural network, it can represent scaling functions more appropriately than polynomials. This method's advantage is that it is computationally less expensive than GP because the computational complexity is linear to the number of data points. With the success of machine learning~\cite{goodfellow2016}, neural networks are being used in many research to solve various problems in physics~\cite{Carleo2019-mu}. Although some studies use machine learning to detect phase transition~\cite{wang2016discovering,ohtsuki2016deep,carrasquilla2017machine}, this paper is the first to use neural networks for FSS to determine critical points and exponents.

We organize this paper as follows. Section~\ref{sec:preliminaries} briefly reviews BSA and neural networks. Section~\ref{sec:NSA} introduces NSA and some techniques for stabilizing the learning process. Section~\ref{sec:numerical_experiment} demonstrates NSA for the two-dimensional Ising model and the two-dimensional bond percolation. Section~\ref{sec:accuracy} introduces a method of calculating the confidence interval of critical values with NSA. Finally, Sec.~\ref{sec:conclusion} gives conclusions and discussion.

\section{Preliminaries}
\label{sec:preliminaries}

\subsection{Bayesian scaling analysis}
Using new rescaled variables,
\begin{align}
    X \equiv (T-\Tc)L^{c_1}, Y \equiv A(T,L)L^{c_2},\label{eq:rescale}
\end{align}
the FSS law is rewritten as follows,
\begin{align}
    Y = F[X].
\end{align}
Thus, the FSS analysis is an inference of critical parameters, $\bm{\theta}_{\mathrm{c}}=(\Tc, c_1, c_2)$, so that rescaled data points, $(\bm{X}_{\bm{\theta}_{\mathrm{c}}}, \bm{Y}_{\bm{\theta}_{\mathrm{c}}})_i = (X_i, Y_i)$, collapse on a scaling function $F$.

The basic idea of the BSA is to assume the scaling function as a sample of the GP and apply the GP regression~\cite{williams2006gaussian}. The scaling function is sampled from the GP as
\begin{align}
    F\sim\mathcal{GP}(0, k_{\bm{\theta}_{\mathrm{h}}}).
\end{align}
It means
\begin{align}
    F[\bm{X}] = (F[X_1], \cdots, F[X_N]) \sim \mathcal{N}(\bm{0}, \Sigma),
\end{align}
where $\mathcal{N}(\bm{0}, \Sigma)$ is a multivariate normal distribution with zero mean vector and the covariance matrix $\Sigma$ as
\begin{align}
    (\Sigma)_{ij} = k_{\bm{\theta}_{\mathrm{h}}}(X_i, X_j),\quad (i=1, \cdots, N)
\end{align}
where $k_{\bm{\theta}_{\mathrm{h}}}$ is a kernel function and $\bm{\theta}_{\mathrm{h}}$ is a hyperparameter vector to define it. Therefore, the BSA regards the rescaled data points in the FSS law as a $N$-dimensional vector of a multivariate normal distribution,
\begin{align}
    \bm{Y}_{\bm{\theta}_{\mathrm{c}}} \sim \mathcal{N}(\bm{0}, \Sigma_{\bm{\theta}}),
\end{align}
where $\bm{\theta} = (\bm{\theta}_{\mathrm{c}}, \bm{\theta}_{\mathrm{h}})$.
Then the conditional probability of data points given parameters $\bm{\theta}=(\bm{\theta}_{\mathrm{h}},\bm{\theta}_{\mathrm{c}})$ is
\begin{align}
    p(\bm{Y}_{\bm{\theta}_{\mathrm{c}}}\mid\bm{\theta})= \frac{1}{\sqrt{\det(2\pi\Sigma_{\bm{\theta}})}}\exp\left[-\frac{1}{2}\bm{Y}_{\bm{\theta}_{\mathrm{c}}}^{\top}\Sigma_{\bm{\theta}}^{-1}\bm{Y}_{\bm{\theta}_{\mathrm{c}}}\right].
\end{align}
Assuming that the prior distribution of parameters $\bm{\theta}$ is uniform for simplicity, we have
\begin{align}
    p(\bm{\theta}\mid\bm{Y}_{\bm{\theta}_{\mathrm{c}}})\propto p(\bm{Y}_{\bm{\theta}_{\mathrm{c}}}\mid\bm{\theta})
\end{align}
from Bayes' theorem.
Therefore, the most probable parameters $\bm{\theta}$ are the maximum of the log-likelihood function,
\begin{align}
    \mathcal{L}(\bm{\theta})= & \log\left[p(\bm{\theta}\mid\bm{Y}_{\bm{\theta}_{\mathrm{c}}})\right]                                                                                                       \\
    =                         & -\frac{1}{2}\log\left[\det(2\pi\Sigma_{\bm{\theta}})\right]-\frac{1}{2}\bm{Y}_{\bm{\theta}_{\mathrm{c}}}^{\top}\Sigma_{\bm{\theta}}^{-1}\bm{Y}_{\bm{\theta}_{\mathrm{c}}}.
\end{align}

One way of finding the maximum of the log-likelihood is the gradient method.
The gradient of the log-likelihood function reads
\begin{align}
    \frac{\partial\mathcal{L}}{\partial\theta}= & -\frac{1}{2}\mathrm{tr}\left(\Sigma_{\bm{\theta}}^{-1}\frac{\partial\Sigma_{\bm{\theta}}}{\partial\theta}\right)
    -\bm{Y}_{\bm{\theta}_{\mathrm{c}}}^{\top}\Sigma_{\bm{\theta}}^{-1}\frac{\partial\bm{Y}_{\bm{\theta}_{\mathrm{c}}}}{\partial\theta}\notag                                                                                                                            \\
                                                & +\frac{1}{2}\left(\Sigma_{\bm{\theta}}^{-1}\bm{Y}_{\bm{\theta}_{\mathrm{c}}}\right)^{\top}\frac{\partial\Sigma_{\bm{\theta}}}{\partial\theta}\left(\Sigma_{\bm{\theta}}^{-1}\bm{Y}_{\bm{\theta}_{\mathrm{c}}}\right).
    \label{eq:gp_grad}
\end{align}
Gradient descent methods, such as Adam~\cite{kingma2017adam}, allow us to obtain the desired critical parameters successfully.
If we take the kernel in the GP as the radial basis function kernel,
\begin{equation}
    k_{\bm{\theta}_{\mathrm{h}}}(X_i, X_j) \equiv {\theta}_{\mathrm{h}, 1}
    \exp\left[
        -\frac{|X_i - X_j|^2}{2({\theta}_{\mathrm{h},2})^2}
        \right],
\end{equation}
sample functions are infinitely differentiable~\cite{kanagawa2018gaussian}.
Therefore, the BSA only assumes the smoothness of the scaling function and hence applies to many systems.
However, the gradient \eqref{eq:gp_grad} includes the inverse calculation of the covariance matrix $\Sigma_{\bm{\theta}}$.
The inverse matrix calculation is generally numerically unstable and has a computational cost of $O(N^3)$.
For this reason, the BSA has the disadvantage of being computationally expensive when the size of data increases.

\subsection{Neural networks}
\label{subsec:nn}
We briefly introduce neural networks and their application to a regression problem, which have played a crucial role in recent machine learning achievements~\cite{goodfellow2016}.

In this paper, we consider a fully-connected neural network with layers numbered from $0$ to $M$, each containing $n_{0},\dots,n_{M}$ neurons.
The network function $\mathsf{NN}\colon\mathbb{R}^{n_{0}}\to\mathbb{R}^{n_{M}}$ can be written with the following recurrence relation:
\begin{align}
     & \alpha^{0}(x) \equiv x,                                  \\
     & \tilde{\alpha}^{l+1}(x) \equiv W^{l}\alpha^{l}(x)+b^{l}, \\
     & \alpha^{l+1}(x) \equiv \phi(\tilde{\alpha}^{l+1}(x)),    \\
     & \mathsf{NN}(x) \equiv \tilde{\alpha}^{M}(x).
\end{align}
Here, $\alpha^{l}$ and $\tilde{\alpha}^{l}$ are functions from $\mathbb{R}^{n_{0}}$ to $\mathbb{R}^{n_{l}}$ for $l=0,1,\dots,M$.
$W^{l}\in\mathbb{R}^{n_{l+1}\times n_{l}}$ and $b^{l}\in\mathbb{R}^{n_{l+1}}$ are weight matrices and bias vectors for $l=0,1,\dots,M-1$.
Each element is a trainable parameter and initialized with $W^{l}_{i,j}\sim\mathcal{N}(0,n_{l}^{-1})$ and $b^{l}_{i}\sim\mathcal{N}(0,1)$.
Total number of training parameters is $\sum_{l=0}^{M-1}n_{l+1}(n_{l}+1)$.
$\phi\colon\mathbb{R}\to\mathbb{R}$ is a nonlinear activation function applied entrywise for a multidimensional input.

Now let us see how we do the regression analysis with neural networks.
We have a training dataset $\mathcal{D}=\{(X_{i},Y_{i})\}_{i=1}^{N}$ with $X_{i}\in\mathbb{R}^{n_{0}}$ and $Y_{i}\in\mathbb{R}^{n_{M}}$,
and we are going to find a function $f$ that satisfies
\begin{align}
    Y_{i} \sim \mathcal{N}[f(X_{i}), E_{i}^2]
\end{align}
with some errors $E_{i}$ for $i=1,2,\dots,N$.
We assume that the function $f\colon\mathbb{R}^{n_{0}}\to\mathbb{R}^{n_{M}}$ can be well approximated by a neural network function $\mathsf{NN}\colon\mathbb{R}^{n_{0}}\to\mathbb{R}^{n_{M}}$, because of
the universal approximation property of neural networks~\cite{Cybenko1989-ii,hornik1991approximation}.
Then, the Gaussian negative log-likelihood loss function is
\begin{align}
    \mathcal{L}=\frac{1}{N}\sum_{i=1}^{N}\frac{1}{2}\left[\frac{\{Y_{i}-\mathsf{NN}(X_{i})\}^{2}}{E_{i}^{2}}+\log (2\pi E_{i}^{2})\right].
    \label{eq:gnll}
\end{align}
We find the desired function by minimizing the loss function by varying parameters in the neural network $\mathsf{NN}$.
Unlike linear regression, we do not have an analytical result on the parameters, but using the stochastic gradient descent method~\cite{ruder2016} allows us to obtain minimizing parameters numerically.
The complicated gradient calculation concerning neural network parameters is done by the automatic differentiation~\cite{baydin2015}, which is implemented in recent deep learning frameworks, including \texttt{JAX}~\cite{jax2018github} and \texttt{PyTorch}~\cite{pytorch2019}.
Also, the computational cost of the loss function \eqref{eq:gnll} is $O(N)$ with respect to the number of data $N$,
which is lighter than the Gaussian process regression.
The comparison between the Gaussian process regression and the neural network approach is made in Appendix~\ref{subsec:calc-time}.

\section{Neural scaling analysis}
\label{sec:NSA}
Let us now introduce the NSA. We explain how NSA is conducted to obtain critical exponents from data. We also give tips on stabilizing the learning process.

\subsection{Data preparation}
\label{subsec:data}
We first prepare the dataset $\mathcal{D}=\{(T_{i},L_{i},A_{i},\delta A_{i})\}_{i=1}^{N}$,
where $A_{i}$ is the $i$th observable data for the temperature $T_{i}$ and the system size $L_{i}$ with a statistical error $\delta A_{i}$.

We next perform the data pre-processing.
In machine learning, data normalization is essential for efficient learning.
We rescale the data as
\begin{align}
     & L\mapsto\frac{1}{L_{\max}}L,                                                                            \\
     & T\mapsto\frac{2}{T_{\max}-T_{\min}}T-\frac{T_{\max}+T_{\min}}{T_{\max}-T_{\min}},\label{eq:T-transform} \\
     & A\mapsto\frac{1}{A_{\max}-A_{\min}}A,\label{eq:A-transform}                                             \\
     & \delta A\mapsto\frac{1}{A_{\max}-A_{\min}}\delta A\label{eq:deltaA-transform},
\end{align}
where $L_{\max}$ is the largest $L$.
$T_{\max},T_{\min}$ and $A_{\max},A_{\min}$ are the largest and the smallest value of $T$ and $A$ for the system size $L_{\max}$, respectively.
Then, as in Eq.~\eqref{eq:T-transform} and Eq.~\eqref{eq:A-transform}, the temperature $T$ is mapped in the interval $[-1,1]$ for $L_{\max}$ and the observable $A$ is mapped so that the width is $1$ for $L_{\max}$.
$\delta A$ is transformed in the same way as $A$ in Eq.~\eqref{eq:deltaA-transform}.
This transformation means that for the scaling function $F$ at the largest system size $L_{\max}$ in Eq.~\eqref{eq:fss},
the data width of the input $(T-\Tc)L^{c_{1}}$ is shaped to be $2$ and the data width of the output $A(T,L)L^{c_{2}}$ is shaped to be $1$.

\subsection{Scaling function}
The basic idea is to replace the scaling function with a neural network.
Since the scaling function is $F\colon\mathbb{R}\to\mathbb{R}$,
we set $n_{0}=n_{M}=1$ for the neural network function $\mathsf{NN}$.
The number of layer $M$ and the layer size $n_{i}$ for $i=1,\dots,M-1$ is left as a hyperparameter, and we set $M=3$ and $n_{1}=n_{2}=20$ in the experiment.

\begin{figure}
    \centering
    \includegraphics[width=\columnwidth]{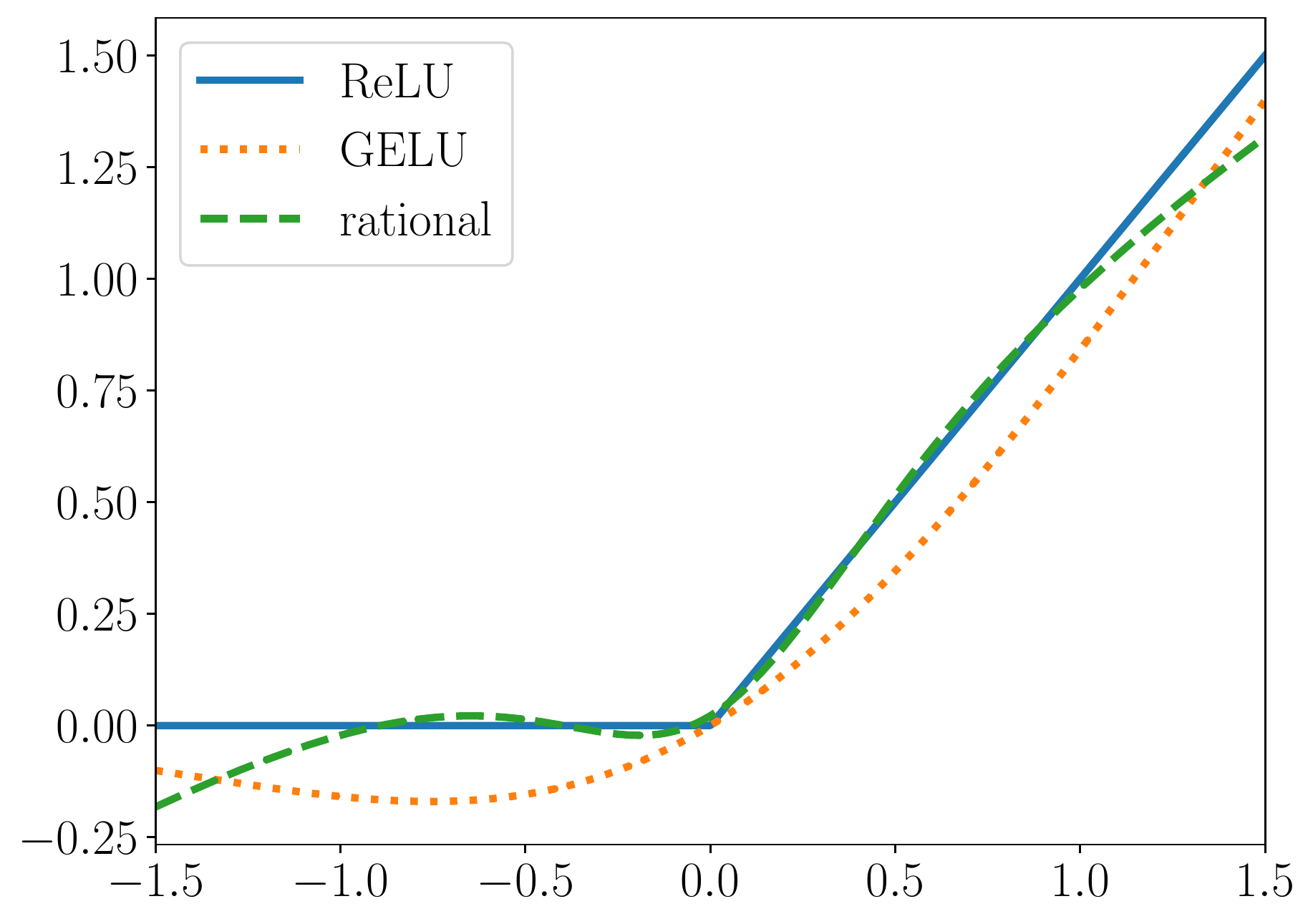}
    \caption{
    Comparison of three activation functions; ReLU function, GELU function~\cite{hendrycks2020gaussian}, and rational activation function with $(r_{P},r_{Q})=(3,2)$~\cite{boulle2020rational}.
    For rational activation function, the parameters $a_{i},b_{j}$ in Eq.~\eqref{eq:rat-act} are initialized
    so that the $L^{\infty}$ distance of ReLU function and rational activation function on $[-1, 1]$ is minimized.
    }
    \label{fig:act-func}
\end{figure}

We remark on the choice of the nonlinear activation function.
In many neural network use cases, the ReLU function, $\mathrm{ReLU}(x)=\max\{x,0\}$, is employed as the activation function to avoid gradient explosion or disappearance.
However, the ReLU function is not differentiable at the origin, which conflicts with our requirement that the scaling function is smooth.
To incorporate this problem, we use the GELU function~\cite{hendrycks2020gaussian},
\begin{align}
    \mathrm{GELU}(x)=x\Phi(x),
\end{align}
which is a smoothed version of the ReLU function.
Here, $\Phi$ is the cumulative distribution function of the standard normal distribution.
We also consider the rational activation function~\cite{boulle2020rational}, which has the form of
\begin{align}
    \phi(x)=\frac{P(x)}{Q(x)}=\frac{\sum_{i=0}^{r_{P}}a_{i}x^{i}}{\sum_{j=0}^{r_{Q}}b_{j}x^{j}},
    \label{eq:rat-act}
\end{align}
where $P(x)$ and $Q(x)$ are polynomials with degree $r_{P}$ and $r_{Q}$.
The coefficients of the polynomials $a_{i},b_{j}$ are trainable parameters.
We use $(r_{P},r_{Q})=(3,2)$ as recommended in the original paper~\cite{boulle2020rational}.
See Fig.~\ref{fig:act-func} for a comparison of the above activation functions.

\subsection{Training the scaling function}
We set up a loss function to learn the scaling function, critical temperature, and critical exponents~\cite{Nix1994}.
Given Eq.~\eqref{eq:fss} for a neural network with Eq.~\eqref{eq:rescale}, we require that the scaling function satisfies the following equation for the data:
\begin{align}
     & Y_{i} \sim \mathcal{N}[\mathsf{NN}(X_{i}), E_{i}^2],
\end{align}
where $X_{i},Y_{i},E_{i}$ are
\begin{align}
     & X_{i}=(T_{i}-\Tc)L_{i}^{c_{1}},                                 \\
     & Y_{i}=L_{i}^{c_{2}}A_{i},\quad E_{i}=L_{i}^{c_{2}}\delta A_{i},
\end{align}
for $i=1,2,\dots,N$.
We remark that the statistical error $E_{i}$ can vary for $X_{i}$
since physical systems that exhibit critical phenomena tend to show large data fluctuation near the critical point.
The Gaussian negative log-likelihood loss function, Eq.~\eqref{eq:gnll},
treats such data with input-dependent noises~\cite{Nix1994}.
In practice, the variances $E_{i}^{2}$ is replaced with $\max(\texttt{eps},E_{i}^{2})$ with a threshold $\texttt{eps}$ for computational stability,
as implemented in \texttt{PyTorch}.
We note that for homoscedastic data where $E_{i}=E$ for all $i$, the Gaussian negative log-likelihood loss function becomes the least-square regression loss function.

The goal of the learning is to find the parameter that gives the minimum value for the loss function $\mathcal{L}$.
We use stochastic gradient descent methods, particularly Adam~\cite{kingma2017adam}, to update parameters and search for optimal solutions.
We summarize the tips obtained through actual experiments.
The learning rate set by default in Adam's training $\alpha=10^{-3}$ is the one recommended for training neural networks.
Therefore, in learning the loss function $\mathcal{L}$ in Eq.~\eqref{eq:gnll},
the learning rate for the parameters of the neural network and the learning rate for the critical exponents $c_{1},c_{2}$ and critical points $\Tc$,
should be different so that the learning can converge in fewer iterations.
In experiments, we set the learning rate for $c_{1},c_{2},\Tc$ as $\alpha_{\mathrm{c}}=10^{-2}$.

We also note how to handle the data. A typical setup for machine learning is batch learning, where the neural network weights are learned using all the data to compute the gradient. However, this method has a problem: if there is a large amount of data, the time required for each learning step becomes enormous. We usually use mini-batch learning to solve this problem, which reduces the time needed for gradient calculation by taking a subset out of the whole data set when calculating the gradient. This method is also well known to make learning less stagnant and optimization more successful. However, we confirmed that batch learning is sufficient for learning in NSA. In all of the following experiments, we used batch learning.

\subsection{Clipping parameters}
In many cases, the sign of the critical exponents $c_{1},c_{2}$ is known in advance when performing the FSS.
Also, by the above rescaling of the temperature $T$, we know that $\Tc$ lies between $-1$ and $1$.
Therefore, we need a way to optimize parameters within an interval or half-open interval.
Here, unconstrained optimization is achieved by constructing a bijective map from the real number $\mathbb{R}$ to the interval.
First, for optimization on the half-open interval $(a,\infty)$, we use the following bijective map:
\begin{align}
    y = g_{(a,\infty)}(x)\equiv\mathsf{softplus}(x)+a,
\end{align}
where $\mathsf{softplus}(x)=\log(1+e^{x})$ is another smooth approximation to the ReLU function.
Second, for the half-open interval $(-\infty, a)$, we use the following bijective map:
\begin{align}
    y = g_{(-\infty,a)}(x)\equiv-\mathsf{softplus}(x)+a.
\end{align}
Finally, the bijective map that achieves parameter optimization on the interval $(a,b)$ is as follows:
\begin{align}
    y = g_{(a,b)}(x) \equiv a + (b-a)\sigma(x),
\end{align}
where $\sigma(x)=1/(1+e^{-x})$ is a sigmoid function.

\begin{figure}
    \centering
    \includegraphics[width=\columnwidth]{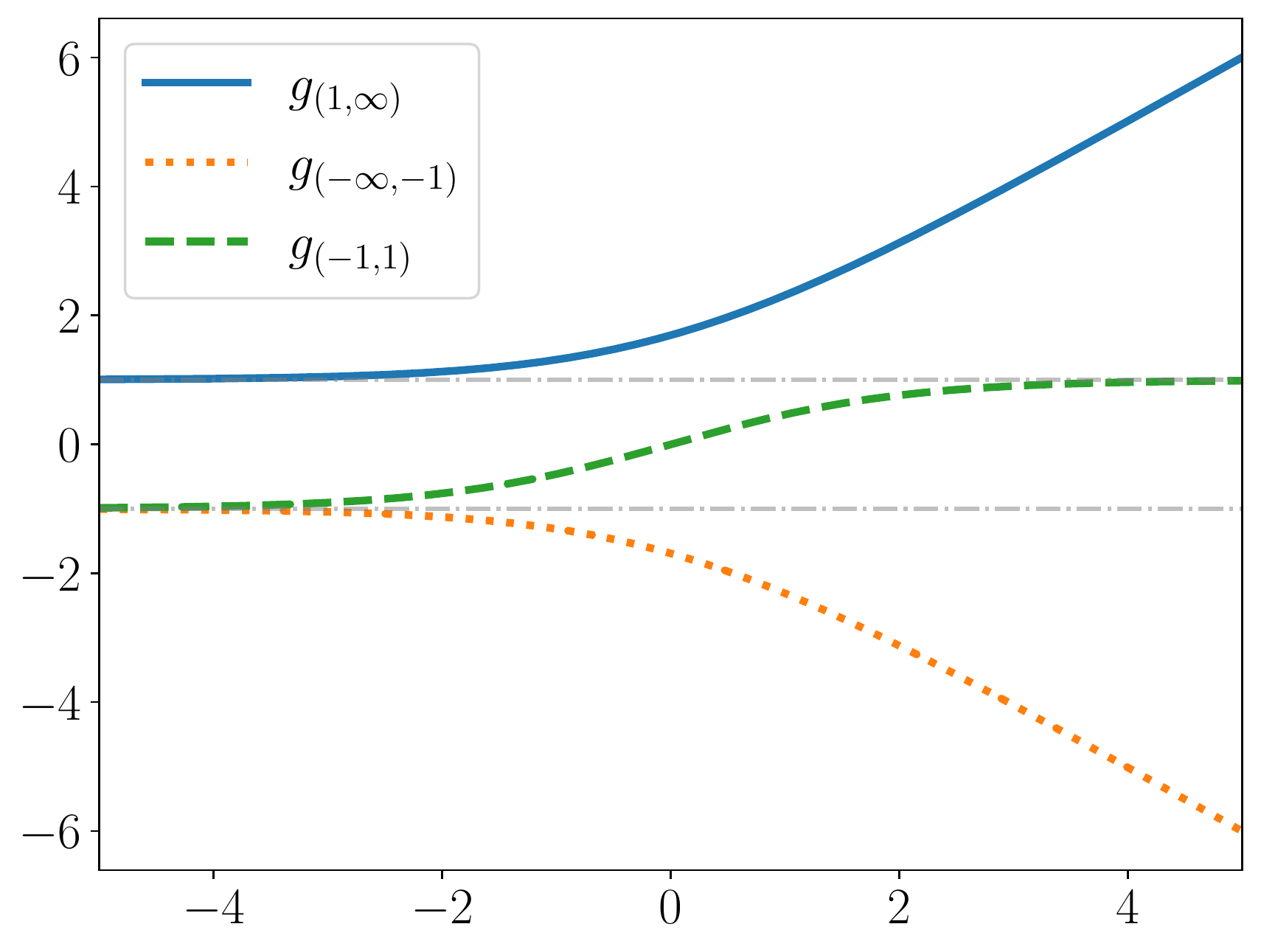}
    \caption{
        Comparison of three bijective maps; $g_{(1,\infty)}$, $g_{(-\infty,-1)}$,
        and $g_{(-1,1)}$. The gray dotted lines are the $1$ and $-1$ guidelines.
    }
    \label{fig:softclip}
\end{figure}

All of these functions $g_{(a,\infty)},g_{(-\infty,a)},g_{(a,b)}$ are bijective maps (see Fig.~\ref{fig:softclip} for the comparison),
and their inverse maps are analytically calculated as follows:
\begin{align}
     & x=g_{(a,\infty)}^{-1}(y)\equiv\log(e^{y-a}-1),  \\
     & x=g_{(-\infty,a)}^{-1}(y)\equiv\log(e^{a-y}-1), \\
     & x=g_{(a,b)}^{-1}(y)\equiv\log(y-a)-\log(b-y).
\end{align}

\section{Numerical experiments}
\label{sec:numerical_experiment}
This section demonstrates NSA for the two-dimensional Ising model and the two-dimensional bond percolation~\footnote{The code implemented in \texttt{Python} is available on \texttt{GitHub} together with the dataset: \texttt{JAX} implementation (\url{https://github.com/yonesuke/jaxfss}) and \texttt{PyTorch} implementation (\url{https://github.com/KenjiHarada/FSS-tools}).}.
In both cases, we can compare the results of NSA with the exact results of those models.
We discuss the performance comparison between NSA and BSA in Appendix~\ref{sec:comparison}.

\subsection{Two-dimensional Ising model}
\label{subsec:ising}
We first consider the two-dimensional Ising model on a square lattice, where the Hamiltonian and its probability distribution are
\begin{align}
     & \mathcal{H}(\bm{s})=-J\sum_{\langle i,j\rangle}s_{i}s_{j},                    \\
     & P(\bm{s})=\frac{1}{Z}\exp\left[-\mathcal{H}(\bm{s})/(k_{B}T)\right].\label{eq:canonical}
\end{align}
Here $s_{i}$ denotes the spin variable of the $i$th site with $s_{i}=\pm1$,
$\langle i,j\rangle$ denotes the nearest neighbor pairs,
and $J$ denotes the positive coupling constant.
$Z$ is the partition function, and $k_{B}$ is the Boltzmann constant.
In the following, we set $J/k_{B}=1$ for simplicity.
The order parameter is a magnetization defined as $M=\sum_{i}s_{i}$.
The two-dimensional Ising model on a square lattice is known to be solvable~\cite{baxter2007},
and the critical point is given by
$\frac{1}{\Tc}=\frac{1}{2}\log(1+\sqrt{2})=0.44068\dots$, and its critical exponents are $\beta=1/8,\gamma=7/4,\nu=1$.

In this paper, we perform FSS using the \textit{Binder ratio} to accurately determine the phase transition point.
The Binder ratio is defined as the kurtosis of the order parameter, which reads
\begin{align}
    U=\frac{1}{2}\left(3-\frac{\langle M^{4}\rangle}{\langle M^{2}\rangle^{2}}\right),
\end{align}
where $\langle\cdot\rangle$ denotes the average over the canonical distribution \eqref{eq:canonical}.
Since the Binder ratio is a dimensionless quantity and does not require scaling,
its FSS form has only one critical exponent $\nu$, which reads
\begin{align}
    U(T,L)=\Psi_{B}[(1/T-1/\Tc)L^{1/\nu}].
\end{align}

\begin{figure}
    \centering
    \includegraphics[width=\columnwidth]{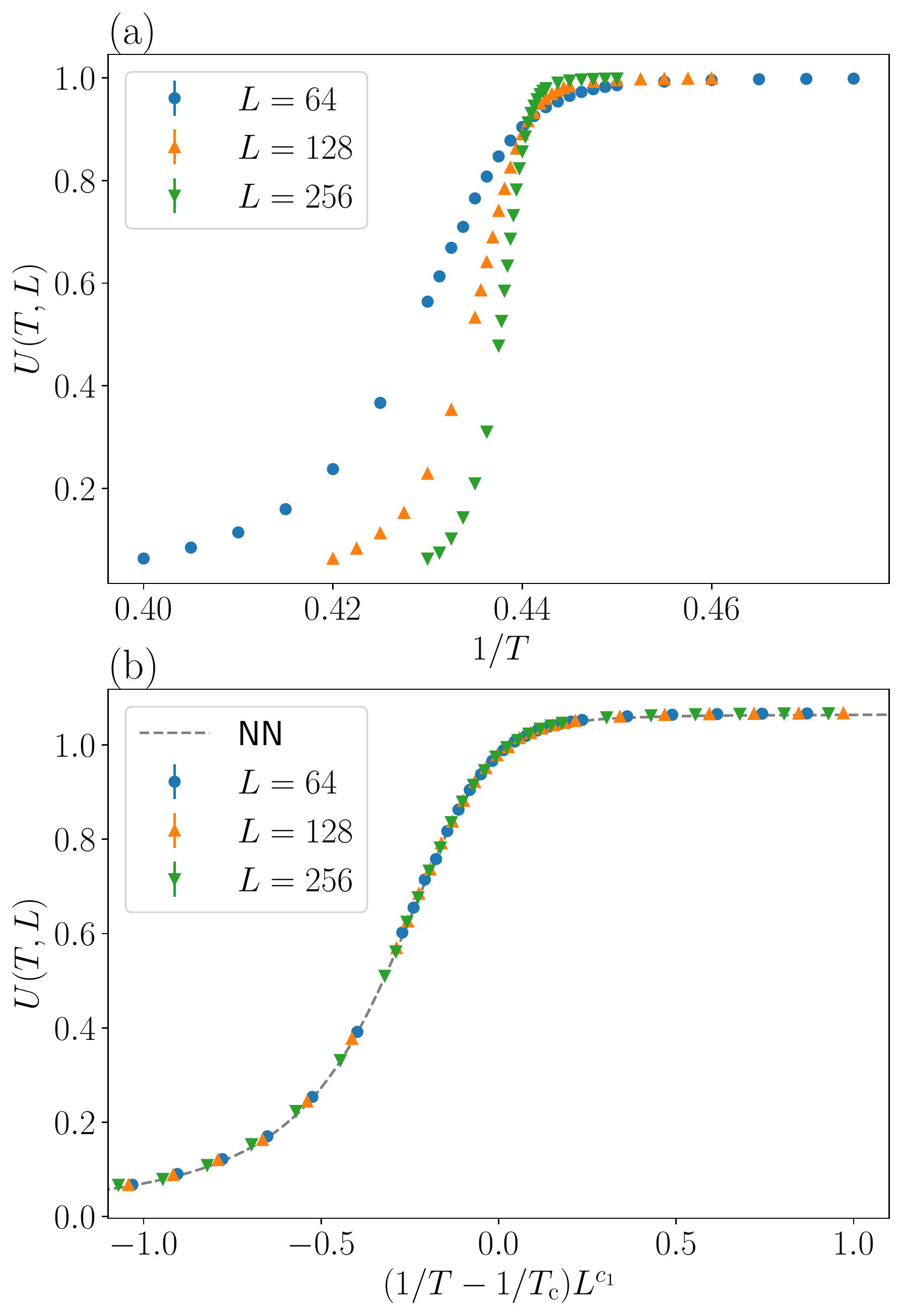}
    \caption{
        (a) The Binder ratios of the Ising model on a square lattice for the system size $L=64,128$ and $256$.
        Error bars are so small that they may not be visible.
        (b) Result of FSS using NSA for the Binder ratios shown in (a).
        The gray dashed curve shows the scaling function obtained by a neural network.
        We note that the data is scaled as in Sec.~\ref{subsec:data}.
    }
    \label{fig:ising_binder}
\end{figure}

We first do the Monte Carlo simulations to obtain the Binder ratio.
See \cite{harada2011} for the detailed settings.
Now we have the dataset $\mathcal{D}=\{(1/T_{i},L_{i},U_{i},\delta U_{i})\}_{i}$, where $U_{i}$ is the $i$th Binder ratio with the inverse temperature $1/T_{i}$ and the system size $L_{i}$.
$\delta U_{i}$ is the statistical error of the $i$th Binder ratio $U_{i}$.
The data is shown in Fig.~\ref{fig:ising_binder} (a) for the system size $L=64,128$ and $256$.
Then to perform the NSA, we rescale the data as in Sec.~\ref{subsec:data}.
The scaling function is set to a neural network with the layer size $n_{0}=n_{3}=1$ and $n_{1}=n_{2}=20$,
and the loss function is set to the Gaussian negative log-likelihood function \eqref{eq:gnll} with
\begin{align}
     & X_{i}=(1/T_{i}-1/\Tc)L_{i}^{c_{1}},   \\
     & Y_{i}=U_{i},\quad E_{i}=\delta U_{i},
\end{align}
where the critical values are clipped with $c_{1}=g_{(0,\infty)}(\theta_{1})$ and $1/\Tc=g_{(-1,1)}(\theta_{\mathrm{c}})$.

\begin{figure}
    \centering
    \includegraphics[width=\columnwidth]{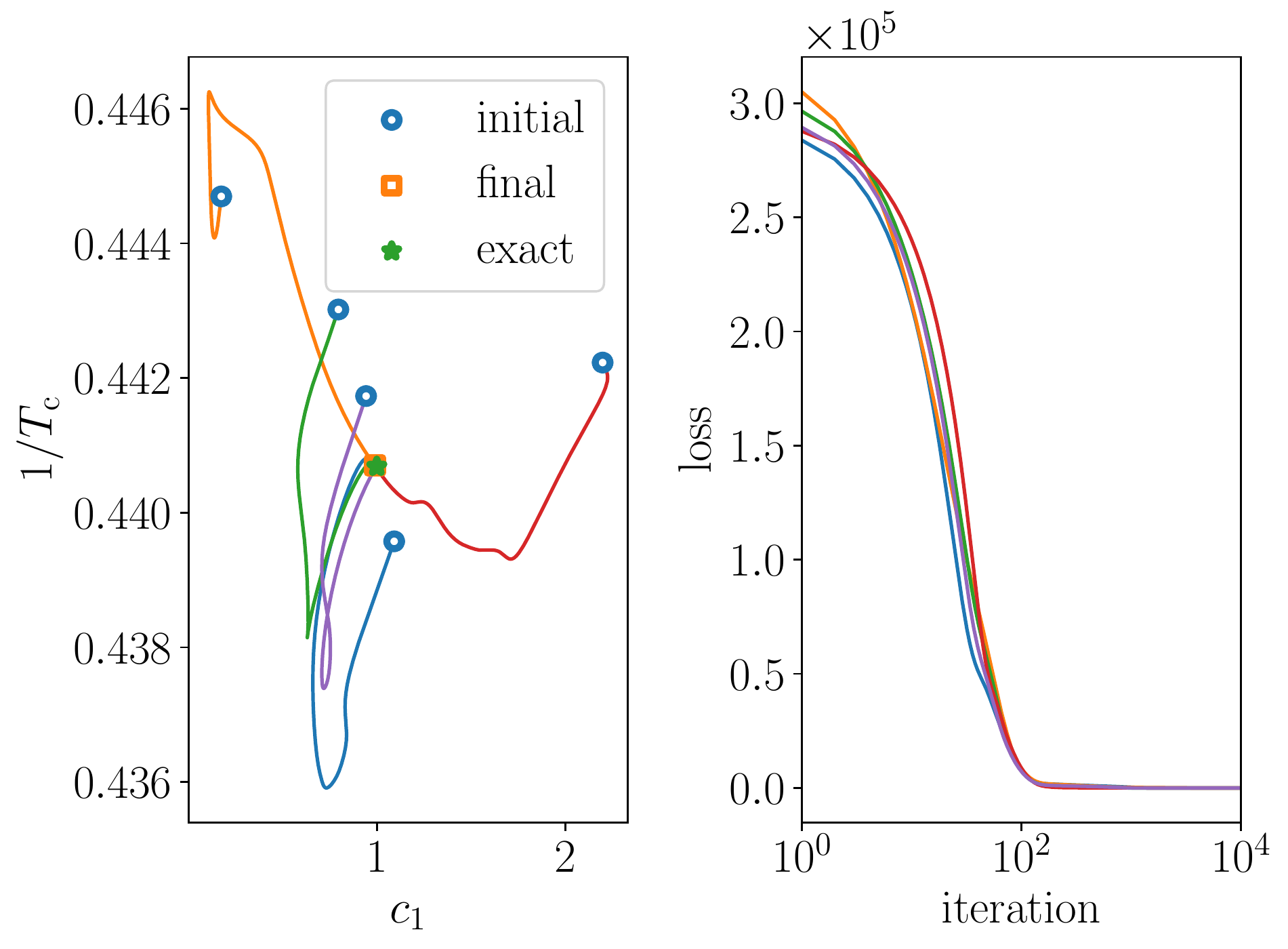}
    \caption{
    Critical values $(c_{1},1/T_{\mathrm{c}})$ (left) and the loss value (right) through the learning process of the Binder ratio of the two-dimensional Ising model with five different initial values.
    }
    \label{fig:ising_loss_params}
\end{figure}

We optimize the scaling function and $\theta_{1},\theta_{\mathrm{c}}$ with the optimizer Adam~\cite{kingma2017adam} for $10^{4}$ iteration times.
Fig.~\ref{fig:ising_binder} (b) shows the result of the NSA.
We see that all the scaled data is on the scaling function obtained by a neural network.
Fig.~\ref{fig:ising_loss_params} shows the loss $\mathcal{L}$ and the critical values through $10^4$ iterations.
We see that the loss values are well converged through iterations.
The results of the critical values are $c_{1}=0.99078$ and $1/\Tc=0.44070$, which are
very close to the exact results.

\subsection{Two-dimensional bond percolation}
\label{subsec:percolation}
\begin{figure}
    \centering
    \includegraphics[width=\columnwidth]{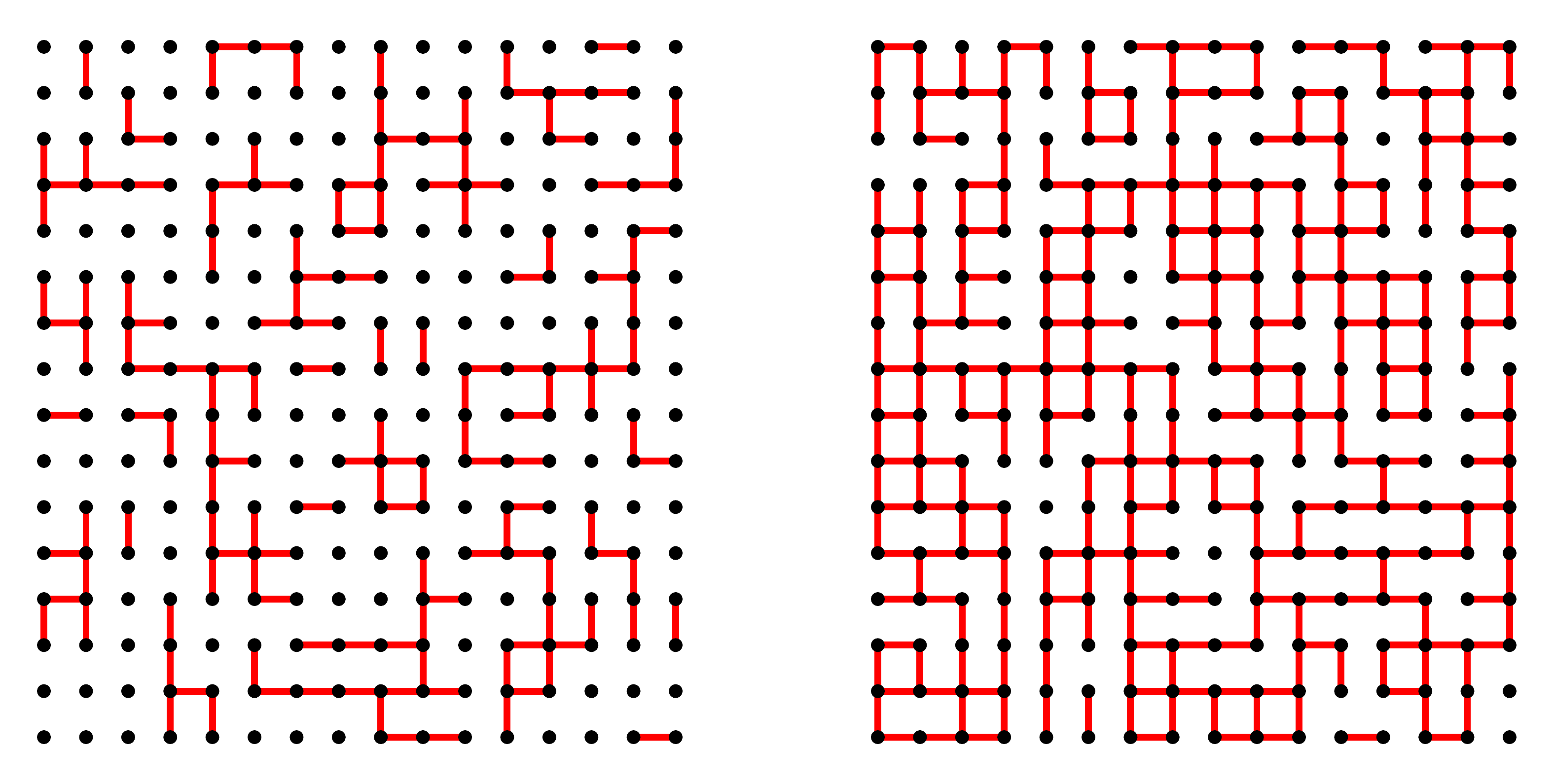}
    \caption{Typical bond percolation configurations on a square lattice with bond probabilities $p=0.3$ (left) and $p=0.6$ (right) for the lattice size $L=16$.
        At $p=0.3$, there are no clusters spread throughout the system.
        When $p=0.6$ exceeds the critical probability $p_{\mathrm{c}}=1/2$,
        clusters spreading throughout the system, i.e., clusters crossing up and down, or left and right, can be confirmed.}
    \label{fig:percolation_configuration}
\end{figure}

Next, we demonstrate NSA for the two-dimensional bond percolation model.
In percolation theory, bonds between lattice points open with probability $p$, then two points are connected.
We discuss the clusters that are then formed, their size, and their dependence on the probability~\cite{Grimmett1999-qx}.
It is known from extensive research to date that as the probability of the opening of bonds between lattice points is increased,
a phase transition phenomenon is observed in which clusters have formed that spread throughout the entire system with a certain probability $p_{\mathrm{c}}$.
The value of this critical probability $p_{\mathrm{c}}$ is known for many two-dimensional lattices.
See Fig.~\ref{fig:percolation_configuration} for typical configurations of two-dimensional bond percolation on a square lattice with a different value of $p$~\cite{Grimmett1999-qx}.

In this paper, we performed FSS using the \textit{truncated mean cluster size} $\chi^{\mathrm{f}}(p)$, which is defined by the mean size of finite open clusters.
By denoting the cluster containing the origin by $C$ and writing its size as $|C|$, we have
\begin{align}
    \chi^{\mathrm{f}}(p)=\mathbb{E}_{p}\left[|C|;|C|<\infty\right].
\end{align}
Near the critical probability $p_{\mathrm{c}}$, the truncated mean cluster size is believed to behave as $\chi^{\mathrm{f}}(p)\approx|p-p_{\mathrm{c}}|^{-\gamma}$
as $p\to p_{\mathrm{c}}$ with a critical exponent $\gamma$.
Therefore its FSS form reads
\begin{align}
    \chi^{\mathrm{f}}(p,L)=L^{\gamma/\nu}\Phi[(p-p_{\mathrm{c}})L^{1/\nu}]
\end{align}
with a scaling function $\Phi$.
For the two-dimensional bond percolation on a square lattice,
the critical probability is $p_{\mathrm{c}}=1/2$, and the critical exponents are $\gamma=43/18$ and $\nu=4/3$~\cite{Grimmett1999-qx}.

We note the difference between the mean cluster size $\chi(p)$ and the truncated mean cluster size $\chi^{\mathrm{f}}(p)$.
The mean cluster size is defined as $\chi(p)=\mathbb{E}_{p}\left[|C|\right]$.
In the subcritical phase $p<p_{\mathrm{c}}$, it is expected that $\chi(p)\approx(p_{\mathrm{c}}-p)^{-\gamma}$ for $p\uparrow p_{\mathrm{c}}$.
Whereas in the supercritical phase $p>p_{\mathrm{c}}$, $\chi(p)=\infty$ from the definition of the critical probability $p_{\mathrm{c}}$,
which makes $\chi(p)$ difficult to handle theoretically.
For the truncated mean cluster size, it is expected that $\chi^{\mathrm{f}}(p)\approx(p-p_{\mathrm{c}})^{-\gamma'}$ for $p\downarrow p_{\mathrm{c}}$, where we assume that $\gamma=\gamma'$.
Since $\chi(p)=\chi^{\mathrm{f}}(p)$ in the subcritical phase,
the formula $\chi^{\mathrm{f}}(p)\approx|p-p_{\mathrm{c}}|^{-\gamma}$ allows us to treat both phases in a unified manner.

\begin{figure}
    \centering
    \includegraphics[width=\columnwidth]{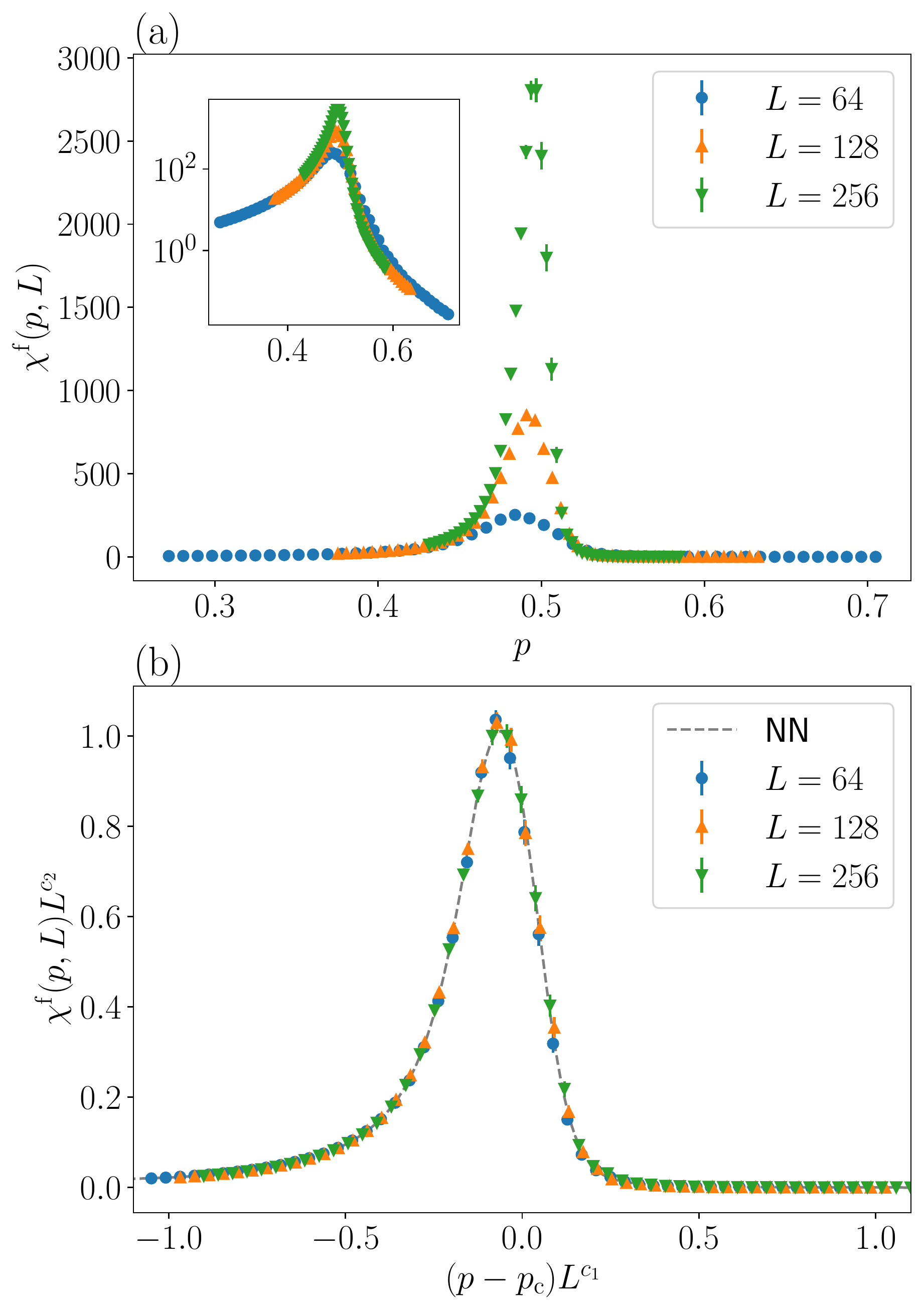}
    \caption{
        (a) The truncated mean cluster sizes $\chi^{\mathrm{f}}(p,L)$ of the bond percolation on a two-dimensional square lattice for the system size $L=64,128$ and $256$.
        Each point is the result of 1000 trials.
        Error bars are so small that they may not be visible.
        The inset shows the $\log$ scale of $\chi^{\mathrm{f}}(p,L)$.
        (b) Result of FSS using NSA for the truncated mean cluster sizes shown in (a).
        The gray dashed curve shows the scaling function obtained by a neural network.
        We note that the data is scaled as in Sec.~\ref{subsec:data}.
    }
    \label{fig:percolation_chif}
\end{figure}

\begin{figure}
    \centering
    \includegraphics[width=\columnwidth]{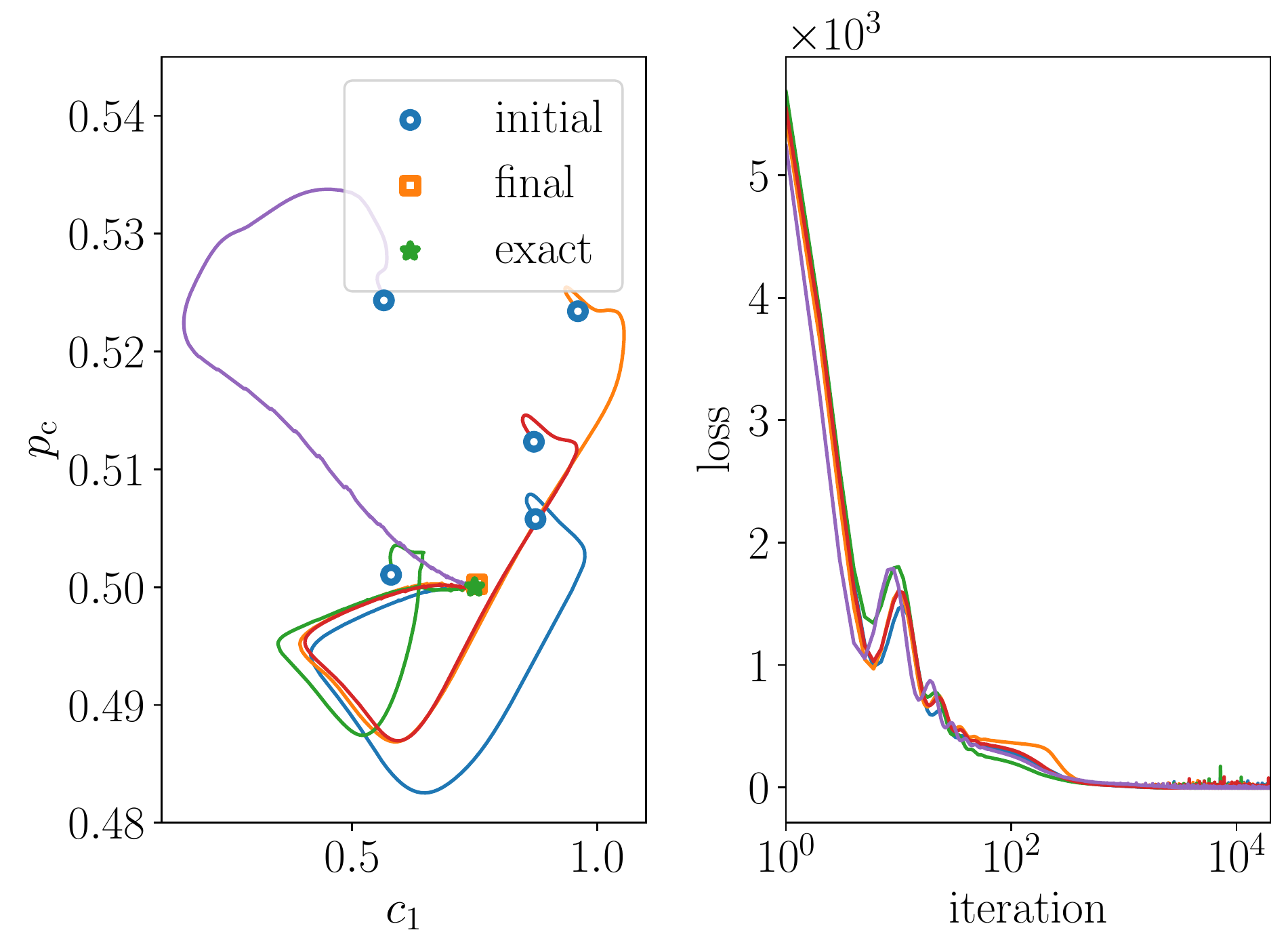}
    \caption{
    Critical values $(c_{1},p_{\mathrm{c}})$ (left) and the loss value (right) through the learning process of the truncated mean cluster size of the two-dimensional bond percolation model with five differential initial values.
    }
    \label{fig:percolation_params}
\end{figure}

To perform FSS, we first do the Monte Carlo simulations to obtain the truncated mean cluster sizes.
Cluster sizes can be calculated efficiently using a \textit{disjoint-set data structure} and the \textit{union-find algorithm}.
We now have the dataset $\mathcal{D}=\{(p_{i},L_{i},\chi^{\mathrm{f}}_{i},\delta\chi^{\mathrm{f}}_{i})\}_{i}$,
where $\chi^{\mathrm{f}}_{i}$ is the $i$th truncated mean cluster size with the probability $p_{i}$ and the system size $L_{i}$.
$\delta\chi^{\mathrm{f}}_{i}$ is the statistical error of $\chi^{\mathrm{f}}_{i}$.
The data is shown in Fig.~\ref{fig:percolation_chif} (a) for the system size $L=64,128$, and $256$ with 1000 trials. The total number of data points is 150.
The rescaling of the dataset is done as in Sec.~\ref{subsec:data}, and the neural network setting is the same as the one with the two-dimensional Ising model in Sec.~\ref{subsec:ising}.
The loss function is set to the Gaussian negative log-likelihood function \eqref{eq:gnll} with
\begin{align}
     & X_{i}=(p_{i}-p_{\mathrm{c}})L_{i}^{c_{1}},                                                     \\
     & Y_{i}=L_{i}^{c_{2}}\chi^{\mathrm{f}}_{i},\quad E_{i}=L_{i}^{c_{2}}\delta\chi^{\mathrm{f}}_{i},
\end{align}
where the critical values are clipped with $c_{1}=g_{(0,\infty)}(\theta_{1})$, $c_{2}=g_{(-\infty,0)}(\theta_{2})$, and $p_{\mathrm{c}}=g_{(-1,1)}(\theta_{\mathrm{c}})$.
The exact values of two-dimensional bond percolation problem are $p_{\mathrm{c}}=1/2=0.5, c_{1}=3/4=0.75,c_{2}=-43/24=-1.7916\dots$. By optimizing the loss function with the Adam optimizer,
we have successfully obtained the critical probability $p_{\mathrm{c}}=0.5002$ and the critical exponents $c_{1}=0.7540,c_{2}=-1.763$. Fig.~\ref{fig:percolation_params} shows the loss $\mathcal{L}$ and the critical values through $10^4$ iterations. In Fig.~\ref{fig:percolation_chif} (b), we see that the scaled data are collapsed on a single curve, and the trained neural network is well approximated to the scaling function.

\section{Accuracy of the estimation of critical values}
\label{sec:accuracy}

In the previous section, we confirmed that the critical values could be estimated using a neural network based on numerical data of critical phenomena. However, this is only one estimation result, and we do not know how much we can trust this result. Most studies that numerically determine the critical value provide confidence intervals for its value. How should we calculate the confidence interval by the NSA proposed in this paper?

Here, we consider using a bootstrap-like approach to calculate confidence intervals for critical values~\cite{efron1992bootstrap}. The bootstrap approach is based on the following procedure:
First, we randomly resample the data of the critical phenomena with replacement.
Next, we take random initial values for the parameters learned in the NSA. The weights of the neural network are initialized randomly as described in Sec.~\ref{subsec:nn}, and the parameters of the critical values $c_{1},c_{2},T_{\mathrm{c}}$ are chosen uniformly at random from the appropriate interval, respectively. Using the dataset and the initial values of the parameters prepared in this way, we obtain the critical values by NSA.
We perform these procedures several times and calculate the standard deviation for a series of critical values, setting this as a $1\sigma$ confidence interval for the critical values.

\begin{figure}[thbp]
    \centering
    \includegraphics[width=\columnwidth]{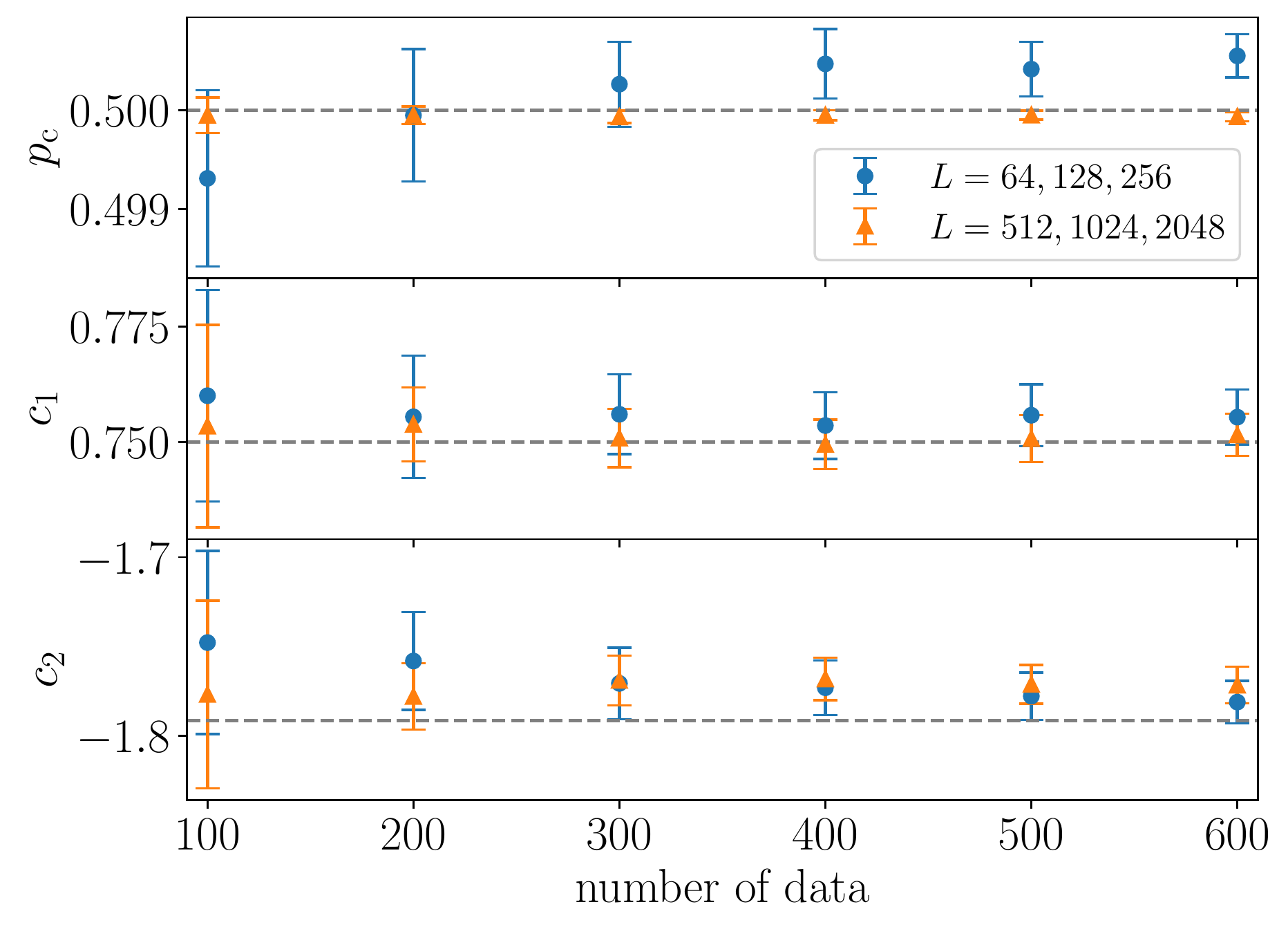}
    \caption{
    Learned critical values $c_{1},c_{2}$ and the critical point $p_{\mathrm{c}}$ for the two-dimensional bond percolation model with confidence intervals indicated by error bars.
    We vary the total number of data from 100 to 600.
    Blue circle points represent the system size triplet $L=64,128,256$,
    and orange triangle points represent the system size triplet $L=512,1024,2048$.
    Confidence intervals are calculated by the bootstrap approach with 500 resamplings.
    True values are plotted with gray dashed lines.
    }
    \label{fig:percolation_acc}
\end{figure}

For example, let us determine the confidence interval of the critical values of the two-dimensional bond percolation model. We use the truncated mean cluster size $\chi^{\mathrm{f}}$ used in Sec.~\ref{subsec:percolation} as our data.
For the initial values of the critical values $p_{\mathrm{c}},c_1$, and $c_2$, we choose uniformly at random from intervals $[0.432,0.583], [0.6,0.9]$, and $[-1.9,-1.6]$, respectively.
Under this configuration, we calculate the critical values 500 times and obtain the confidence interval.
In Fig.~\ref{fig:percolation_acc}, we use the truncated mean cluster size for the data with system size triplets $L=64,128,256$ and $L=512,1024,2048$.
We vary the total data size from 100 to 600 and plot the learned critical values with $1\sigma$ confidence intervals.
As can be seen in the figure, the error bar becomes smaller as the data size increases,
indicating that we get a more accurate result for a larger number of data for the two-dimensional percolation model.
We also observe that for $p_{\mathrm{c}}$ and $c_{1}$, the accuracy with respect to the true value improves as the system size $L$ of the data used for training is increased.

\section{Conclusions and discussion}
\label{sec:conclusion}
We proposed a method for the scaling analysis of critical phenomena using neural networks. The basic idea of this method is to model the scaling function by a neural network. For example, we consider FSS analysis, but we can apply it to general scaling analysis, which contains an unknown scaling function. The method is computationally lighter than the previously proposed method for scaling analysis using GP regression, making it a simpler method.

We demonstrate it for the two-dimensional Ising and bond percolation models. We could accurately obtain the critical points and exponents, and the data are well-collapsed to the learned scaling function for both cases.
Because the computational complexity is linear for the number of data points, we can handle the FSS analysis efficiently.

Using a bootstrap-like approach, we also check the robustness of the estimation results of the new method and define the confidential intervals as the standard deviation of estimated values. They are consistent with the exact values. Another approach to determine the accuracy of the critical values would be to use the \textit{stochastic gradient Langevin dynamics}~\cite{welling2011bayesian}, which achieves the Bayesian learning by adding noise to the gradient method. It has the advantage of being a Bayesian approach but scales linearly in computational complexity for the number of data.

\acknowledgments
The authors thank the anonymous referees for their valuable comments and for making the manuscript a better one.
RY acknowledges the support of JSPS KAKENHI Grant Number JP22J15552.
KH acknowledges the support of JSPS KAKENHI Grant Number JP20K03766 and a Grant-in-Aid for Transformative Research Areas ``The Natural Laws of Extreme Universe -- A New Paradigm for Spacetime and Matter from Quantum Information'' (KAKENHI Grant Numbers JP21H05182, JP21H05191) from JSPS of Japan.

\appendix
\section{Comparison between NSA and BSA}
\label{sec:comparison}

In this section, we compare the performance of NSA and BSA, particularly with respect to computation time and accuracy of critical values.

\subsection{Computation time}
\label{subsec:calc-time}
\begin{figure}
    \centering
    \includegraphics[width=\columnwidth]{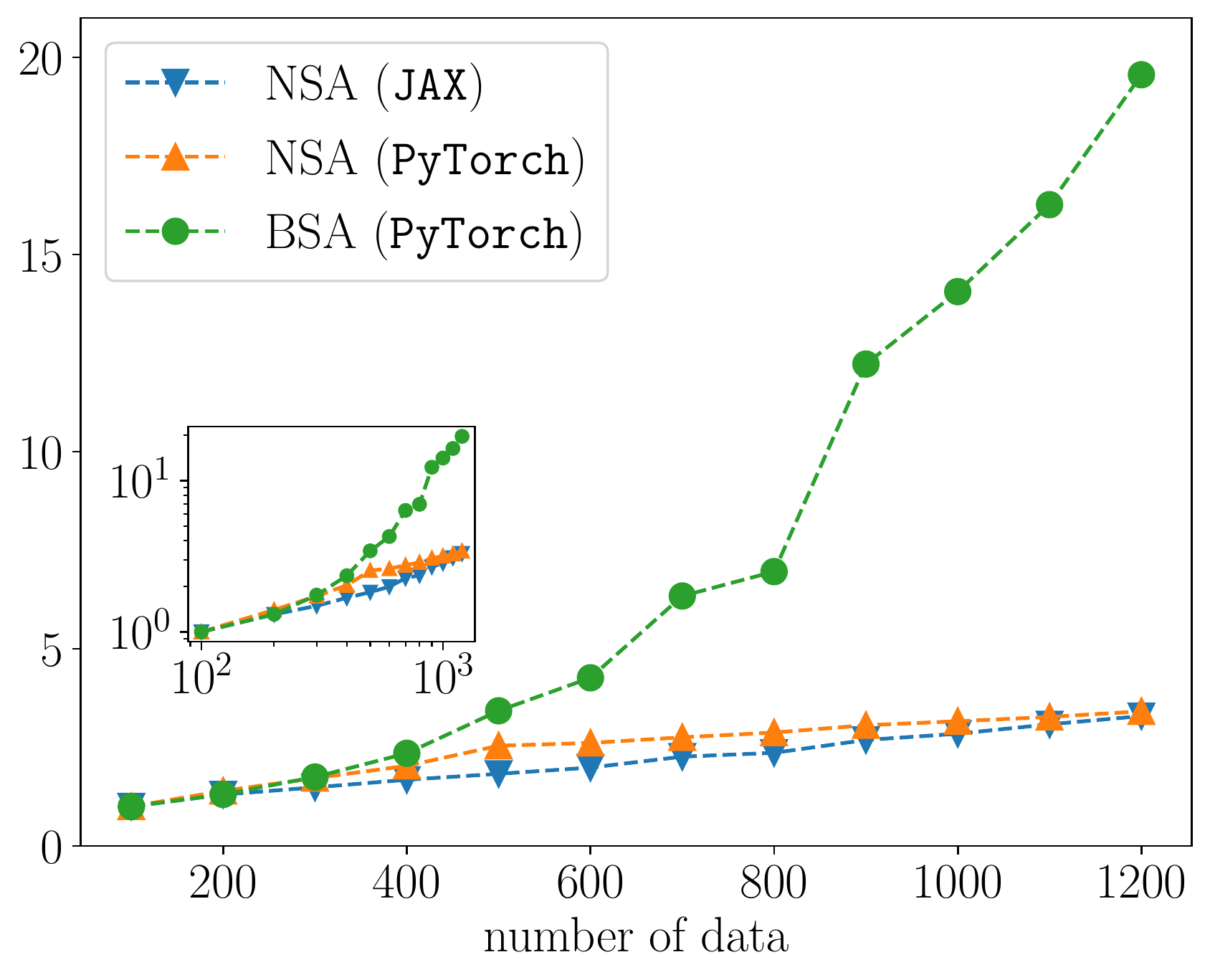}
    \caption{Relative average execution time for one epoch where we scale the execution time for 100 data as 1.
    The data used is the truncated mean cluster size of the two-dimensional bond percolation.
    The inset shows the log-log plot.}
    \label{fig:time}
\end{figure}

Let us compare the execution time by NSA with BSA with respect to the number of data.
\texttt{PyTorch} and \texttt{JAX} implementations were used to execute NSA.
\texttt{PyTorch} implementation was used to execute BSA.
The execution environment is as follows: [OS] macOS 12.5.1, [CPU] Apple M1 Max, [Memory] 32GB.
In Fig.~\ref{fig:time}, we measured the (average) computation time per epoch for NSA and BSA and plotted how much the computation time increases as the number of data increases, taking the case with 100 data as 1.
As can be seen, the computation time of NSA is linear to the number of data,
while BSA shows a larger growth rate of computation time with respect to the number of data.
This is because BSA requires $O(N^{3})$ computational complexity due to inverse matrix calculation,
while NSA requires only linear $O(N)$ computational complexity.
We note that the \texttt{JAX} implementation uses \texttt{jax.jit()}, which performs \textit{just in time} (JIT) compilation~\cite{jax2022jit,tf2022xla},
so it allows for faster execution than \texttt{PyTorch}.

\subsection{Accuracy of critical values}

\begin{figure}[thbp]
    \centering
    \includegraphics[width=\columnwidth]{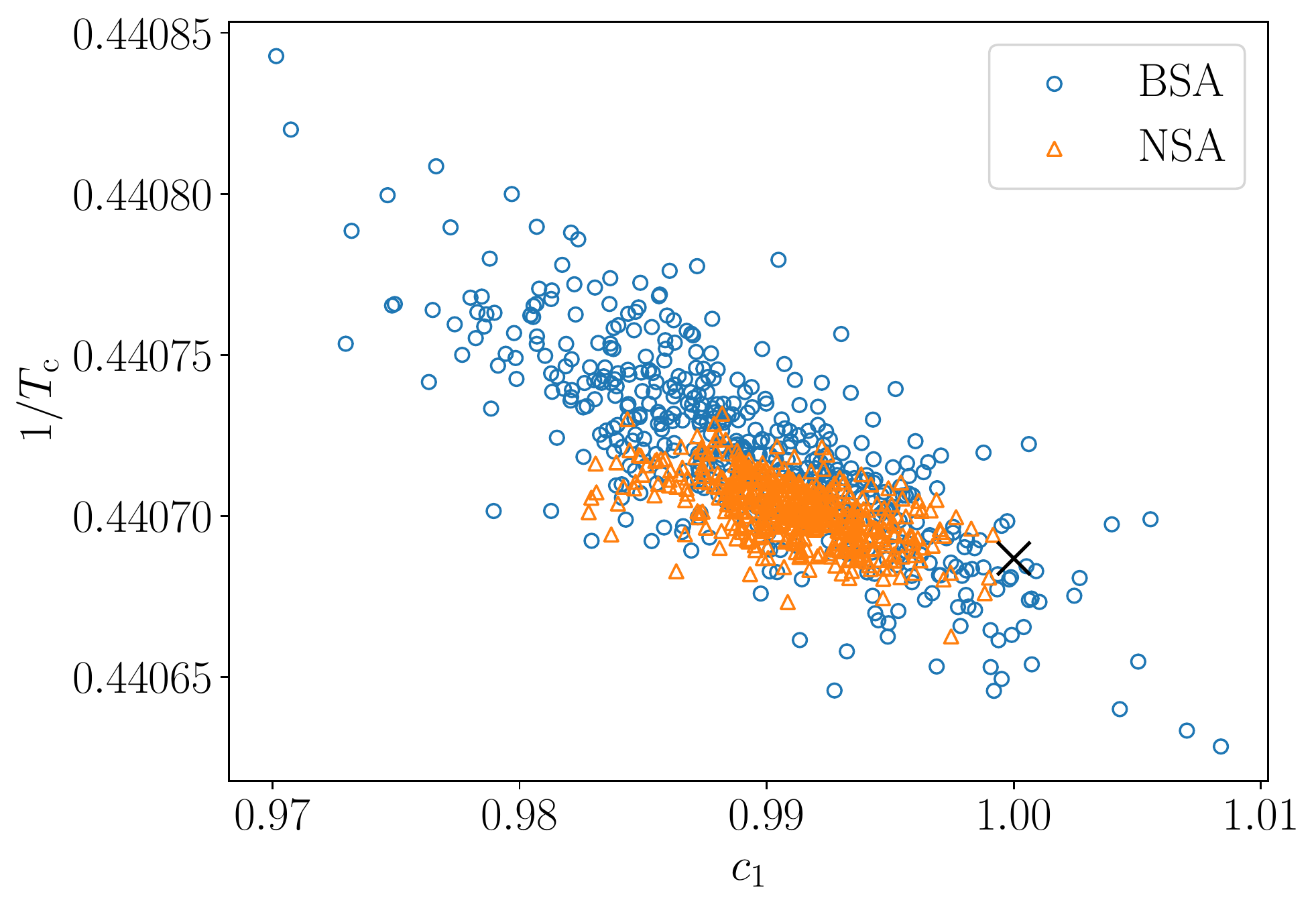}
    \caption{Final learned value distributions of critical values $(c_{1},1/T_{\mathrm{c}})$ for NSA and BSA for two-dimensional Ising model.
    The cross mark represents the theoretical value $c_{1}=1$ and $1/T_{\mathrm{c}}=\frac{1}{2}\log(1+\sqrt{2})$
    }
    \label{fig:ising_nsa_bsa}
\end{figure}

We next calculate the confidence intervals of critical values for the two-dimensional Ising model
and compare the accuracy of NSA and BSA methods.
For data used in Sec.~\ref{subsec:ising}, we resample the data with replacement and calculate the critical values for 500 times.
Figure~\ref{fig:ising_nsa_bsa} shows the results of bootstrap for NSA and BSA.
We observe that the learned results are more concentrated in the NSA compared to the BSA.
For the BSA method, $c_{1}=0.989(5)$ and $1/T_{\mathrm{c}}=0.44071(3)$,
whereas for the NSA method, $c_{1}=0.991(2)$ and $1/T_{\mathrm{c}}=0.44070(1)$.

\bibliography{main}

\end{document}